\documentstyle{l-aa}

\begin{document}

\thesaurus{3(11.16.1;11.19.6;11.09.4;13.09.1)}

\title{Near-infrared observations of galaxies in Pisces-Perseus:
II. Extinction effects and disk opacity}

\author{ G. Moriondo \inst{1,2}
\and
R. Giovanelli \inst{2}
\and
M.P. Haynes \inst{2}
}
\institute{ 
  Dipartimento di Astronomia e Scienza dello spazio, Universita' di Firenze,
L. E. Fermi 5, I-50125 Firenze, Italy
 \and  
  Center for Radiophysics and Space Research, Cornell University, Ithaca, NY 
14853 
} 

\offprints{G.~Moriondo}
\date{Received ; accepted }

\maketitle
\markboth{Moriondo et al.: Extinction in spiral galaxies}{Moriondo: Extinction 
in spiral galaxies}

\begin{abstract}

We study the correlations with inclination of $H$-band disk and bulge 
structural 
parameters and $I-H$ colour profiles for a sample of 154 spiral galaxies,
in order to detect possible effects due to internal extinction by dust.
The selection of the sample assures that galaxies at different 
inclinations are not intrinsically different, so that the observed 
correlations represent the real behaviour of the parameter considered.
All the parameters are derived from a bi-dimensional fitting of the 
galaxy image. We find that extinction, though small at near infrared 
wavelengths, is sufficient to produce observable effects. In particular the
observed increase of the average disk scalelength and the reddening of the 
disk $I-H$ colour at high inclinations are clear signatures of the presence 
of dust. The total $H$-band disk luminosity depends little on inclination;
on the other hand significant corrections to the face-on aspect are 
derived for the $H$-band central disk brightness and the disk scalelength. 
The bulge parameters exhibit little or no dependence on inclination.
Simulations carried out with a simple model for an internally-extincted 
galaxy show that these results imply a central $H$-band optical depth 
between 0.3 and 0.5.
\keywords{Galaxies: photometry -- structure -- ISM -- Infrared: galaxies}
\end{abstract}

\section{Introduction}

The amount of internal extinction in spiral disks has been the subject 
of a long series of papers throughout the last few years, since Disney et al.
(\cite{disney}, DDP hereafter) and Valentijn (\cite{vale}), questioning the 
widespread assumption of transparency of galaxies, suggested that disks might
actually be optically thicker than previously thought. In particular their 
claim was in disagreement with the conclusions by Holmberg dating back to 
\cite{holm1} and \cite{holm2}.
Subsequent papers (e.g. Peletier \& Willner \cite{pw}; Jansen et al. 
\cite{jansen}; Giovanelli et. al \cite{g94} -- G94 hereafter) tend to support a 
scenario in which galaxies are transparent in the outer regions and moderately 
opaque at the center. Though this is likely to be the case, the actual 
amount of extinction in spiral disks is still a matter of debate.
On the other hand, estimating the internal extinction in disks and how it 
affects the observed light distribution is necessary in order to correct 
the observed quantities to either the face-on aspect or the dust-free case.
Such corrections are 
fundamental in any study concerning galaxies both as single objects (structure 
and evolution), and as tracers of the large scale 
structure of the universe.

A possible approach to characterizing the extinction properties of disks
relies on the statistical properties of large samples of galaxies, for 
example, analyzing the correlations of observable parameters with inclination.
In this way, average corrections to the face-on aspect for such quantities 
can be calculated in principle without introducing any modeling of the 
extinction processes. On the other hand, simulations of extincted galaxies 
(e.g. Byun et al. \cite{byun}; Bianchi et al. \cite{bianchi}; Corradi et al.
\cite{corradi}) can predict 
the trends observed, and therefore provide insights into the parameters of 
the dust 
layer, such as the thickness or the central optical depth $\tau(0)$.
A few caveats need to be considered when undertaking a study of this 
kind. First of all, simulations have shown that some tests are not
significant at any optical depth, since
in some cases a large variation in $\tau(0)$ has only a minor influence on 
a particular parameter. Using many different correlations at the same time 
will necessarily yield a more confident estimate of the opacity.
Of course every structural parameter (scalelength, surface 
brightness, the inclination itself) needs to be determined in a 
most reliable and objective way, possibly separating the two contributions
from bulge and disk to the total brightness distribution.  
If simulations are used to predict the observed trends with inclination,
a comparison is possible only if real and simulated data are analyzed in a 
similar way. In fact different decomposition techniques 
often yield discrepant estimates of the measured parameters as pointed out,
e.g., by Knapen \& van der Kruit (\cite{knapen}).
Finally, in the galaxy sample used for the analysis, possible observational 
biases need to be accounted for, as demonstrated by several authors 
(Burstein et al. \cite{bhf}; Davies et al. \cite{dpbd}; Giovanelli et al. 
\cite{g95} -- G95 hereafter).
Davies et al., for example, showed that galaxies from a 
diameter-limited sample tend to be characterized by a preferred value of the 
central disk surface brightness $\mu(0)$, which is determined by the 
visibility function of that sample (Davies \cite{davies}). 
This preferred value does not depend on the inclination or on the intrinsic 
distribution of surface brightness, so that any conclusion about the disk 
opacity derived by simply investigating the observed correlation of
$\mu (0)$ with inclination is in general affected by this selection effect. 

In this paper, we study how the bulge and disk structural parameters correlate 
with inclination in the near infrared (NIR), in order to detect and 
quantify the effect of internal extinction at these wavelengths. 
To this purpose we use a sample of 154 spiral galaxies in
the Pisces-Perseus supercluster region, for which $H$ band images are 
available. These data were collected for a project aimed to investigate
the structural, photometric and dynamic properties in the NIR of a large
homogeneous sample of spirals, for which a reliable estimate 
of the amplitude of internal extinction is necessary. This study 
can also provide information on the general issue of disk opacity,
beyond its NIR perspective.
In fact, if disks are optically thick in the $B$ band, 
we expect them to be only partly opaque in the NIR, and simulations
clearly show that some of the correlations with inclination (in particular
the correlations of central brightness and scalelengths) are useful
and significant only in the case of {\it moderate} opacity.

The separation of the bulge and disk contributions to the total light, and
the subsequent determination of the various structural parameters of 
the two components, is carried out using the entire galaxy image, instead of
a brightness profile, which considerably improves the reliability of the 
results (Byun \& Freeman \cite{bf}; Moriondo et al. \cite{moriondo}).
Since the sample is selected from a single supercluster the scatter introduced 
by distance selection biases are minimized.
For about 70 galaxies of the sample $I$ band brightness profiles are also
available, from the ScI sample of Haynes et al. (in preparation).
for these objects $I-H$ colour profiles were also used to investigate the 
extinction properties of the disks.

We first describe the data analysis and discuss the
correlations between the various parameters and the inclination, summarizing
briefly the trends expected at different opacity levels. Next we account 
for possible selection effects on the sample and define an
intrinsically homogeneous set of galaxies. We finally discuss the trends 
observed in our sample and we compare our results with the 
predictions of a model for an extincted galaxy, deriving an estimate 
for the central optical depth in the $H$ band.
Throughout the paper we assume a Hubble constant 
$H_0=75$~km~s$^{-1}$~Mpc$^{-1}$.
	
\section{The Sample}

Our galaxies belong to a sample selected in the region of the Pisces-Perseus
supercluster and were initially drawn from the target of the redshift survey
of this region carried out by R.G. and M.H. 
Selection criteria, observations, data reduction and photometric calibration 
are described in Moriondo et al. (in preparation).
One hundred seventy eight galaxies were imaged in the NIR photometric bands
($J$, $H$, and $K$), with types ranging from S0 to Sd and covering the whole 
range of inclinations -- though with fewer objects at high inclination. 
Since for a large fraction of these objects only $H$ band data are available, 
in the remainder of the paper we will consider only the $H$ band subsample 
(174 galaxies as a whole, mostly late-type spirals).

Distances were evaluated correcting 
the heliocentric radial velocities to the CMB frame of reference with 
the prescription reported in RC3 (de Vaucouleurs et al. \cite{rc3}).
The average distance of the sample is 70 Mpc with a dispersion of 24,
yielding an average $H$-band absolute magnitude of $-23.4$; the luminosity
range is between $\sim -21$ and $\sim -25$ $H$-mag.

\section{Data analysis}

A surface brightness profile was extracted for each galaxy using the ELLIPSE 
routine in the STSDAS\footnote{STSDAS is distributed by the Space
Telescope Science Institute, which is operated by the Association of
Universities for Research in Astronomy (AURA), Inc., under NASA contract
NAS 5--26555.} package for data reduction and analysis. In particular, the 
ellipse fitting was performed keeping the center 
of the ellipses fixed, but allowing ellipticity and position angle
to vary.
From such profiles we determined the isophotal radii considered
in the following analysis. They were also used to obtain $I - H$ colour 
profiles, as described in Sect. \ref{sec:colour}.

To obtain a description as reliable and objective as possible of 
the galaxy structure, rather than merely determining the outer slope 
of the major axis profiles, we fitted a bi-dimensional model 
brightness distribution convolved with a gaussian PSF to each galaxy image. 
The model includes two components, a disk and a bulge, whose
apparent brightness distributions are both elliptical and decrease
exponentially with radius. In a magnitude scale, along the major axis:
\begin{equation}
\protect\label{eq:bd}
\mu (r) = \mu (0) + 1.086 \frac{r}{r_d}  \;\; . 
\end{equation}
We expect an exponential shape for the bulge to 
be appropriate for these galaxies since almost all of them are
Sb's or later types (see for example Andredakis et al. \cite{apb}), and in
fact this choice is supported a posteriori by the good quality achieved by 
most of the fits.
When the exponential distribution for the bulge is too steep
to yield a satisfactory fit to the data (7 cases out of 174) we consider a 
generalized distribution for this component (S\` ersic \cite{sersic}), i.e.: 
\begin{equation}
\label{bulge_b}\protect
\mu_b\,(r)\ = \mu_e + 1.086
\left\{-\alpha_n \left[\left(\frac{r}{r_e}\right)^{1/n}-1\right]\right\}\;\;.
\end{equation}
The exponent index $n$ -- equal to 1 for the exponential distribution -- 
is allowed to be an integer greater than 1; $\mu_e$ and $r_e$ are the effective 
surface brightness and radius, and $\alpha_n$ is a constant relating the 
effective brightness and radius to 
the exponential values (see Moriondo et al. \cite{moriondo}).
In the cases for which exponential bulge fits are unsuccessful a good fit 
to the data  is obtained with $n=2$.
The parameters of each fit are the two scalelengths, the two surface 
brightnesses, and the two apparent ellipticities of bulge and disk. 
We use the fitted disk ellipticities to determine the axial ratios $a/b$ used 
for the correlations in the next sections.
For 8 galaxies with particularly disturbed morphology we do not manage
to obtain a satisfactory fit; these objects will therefore be excluded from 
the following analysis.
More details about the fitting procedure can be found in Moriondo et al.
\cite{moriondo}.

\section{Basic relations between disk parameters and inclination}

Both the central surface brightness of the disk, $\mu(0)$, and its scalelength
$r_d$, are expected to correlate with inclination in a way which 
depends on the amount of extinction present in the disk. 
This is also true for isophotal radii and total luminosities, which 
can be expressed as functions of $\mu(0)$ and $r_d$. From Eq. (\ref{eq:bd}) 
a generic isophotal radius $r_{iso}$ is defined by:
\begin{equation}
\protect\label{eq:riso_def}
r_{iso} = 0.921 (\mu_{iso} - \mu (0)) r_d \;\;  ; 
\end{equation}
whereas the total magnitude of the disk is given by
\begin{equation}
\protect\label{eq:m_def}
m = \mu(0) - 2.5 \log (2\, \pi r_d^2) + 2.5 \log (a/b) \;\; .
\end{equation}
If the galaxy is completely transparent $\mu(r)$ is expected to become 
brighter with inclination:
\begin{equation} 
\protect\label{eq:cdbi}
\mu(r) = \mu^{\circ}(r) - 2.5 \log (a/b), 
\end{equation}
where we have denoted the face-on value with a ``$^{\circ}$'';
in this case Eq. \ref{eq:bd} implies that the disk scalelength will be 
independent of $a/b$, and from Eq. (\ref{eq:m_def}) it follows that $m$ 
is also independent of inclination,  whereas from Eq. (\ref{eq:riso_def}) 
we obtain:
\begin{equation}
\protect\label{eq:c_ri_in}
r_{iso} = r_{iso}^{\circ} + 2.306\, r_d \log(a/b)
\end{equation}

If some extinction is present in the disk Eq. (\ref{eq:cdbi}) can be 
generalized as
\begin{equation}
\protect\label{eq:cm0i}
\mu (0) = \mu^{\circ} (0) -2.5\, C\log (a/b) \;\; ,
\end{equation}
where $0 \leq C \leq 1$, with $C=0$ corresponding to the completely 
opaque case. Following G94, relationships of other photometric parameters
with inclination are also parameterized as linear dependences on $\log (a/b)$.
For the disk scalelength, the isophotal radius and the total 
disk magnitude we will have respectively:
\begin{eqnarray}
r_{d}   & = & r_{d}^{\circ}\, [1+\eta \log(a/b)]  \protect\label{eq:c_r_in}  \\
r_{iso} & = & r_{iso}^{\circ}\, [1+\delta_{iso} \log(a/b)] \protect\label{eq:c_riso_in}  \\
m_{d} & = & m_{d}^{\circ}+\gamma \log(a/b) \;\; . \protect\label{eq:c_m_in}    
\end{eqnarray}

Disk scalelengths are expected to be independent of inclination
($\eta = 0$) both in the case of completely transparent and completely 
opaque disks while they will increase with inclination if some extinction is
present in the inner part of the disks (Byun et al. \cite{byun}; Bianchi
et al. \cite{bianchi}).
As opacity increases, any isophotal radius will 
become less sensitive to inclination than in the transparent case 
($\delta_{iso}$ will decrease), whereas we expect the total luminosity to 
become a steeper function of inclination ($\gamma$ will increase).
We will assume that relations analogous to Eqs. \ref{eq:c_r_in}-\ref{eq:c_m_in}
are appropriate to represent the trends of the corresponding bulge 
parameters as well. 

The relations considered so far hold both for apparent and absolute 
quantities if we consider a single galaxy. Yet, since we are looking for 
average correlations in an extended sample, we will use only 
distance-independent quantities, i.e. the absolute ones, for which we 
will maintain a notation analogous to that shown above.

\section{Checking for selection effects \protect\label{sec:sel}}

To check for possible intrinsic differences between high
and low inclination galaxies in our sample we assume 
that at $H$ band galaxy disks are transparent at all inclinations at 
sufficiently large isophotal radii. In particular, we choose the radius 
where the surface brightness equals 21 $H$ mag arcsec$^{-2}$, which 
corresponds on average to a distance 
of three disk scalelengths from the center and to the 22.5 mag isophote
in the $I$ band. We note that even with a $B$ band face-on
optical depth as high as 20, the triplex model by DDP
and a standard extinction curve (e.g. Cardelli et al. \cite{cardelli})
predict an optical depth as low as 0.1 in the $H$ band at about three disk
scalelengths from the center. More refined models and simulations (e.g.
Byun et al. \cite{byun}) confirm that the variation of the outer isophotal
radii with inclination is largely independent of $\tau(0)$. 
In the hypothesis of transparency, we can estimate the face-on value of 
$r_{21}$ from Eq. (\ref{eq:c_ri_in}) for every galaxy in the sample.
A fair sample should cover the same range in $r_{21}^{\circ}$ at all 
inclinations.

\begin{figure*}
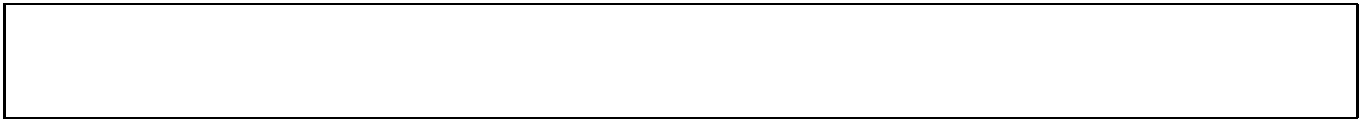

\picplace{1.5cm}
\caption[]{Face-on corrected isophotal radius at 21 $H$ mag arcsec$^{-2}$ 
versus the logarithm of inclination. The galaxies selected for the following
analysis are the ones between the two horizontal dotted lines (respectively
at 2.9 and 16.5 kpc). Crosses correspond to Sa galaxies and earlier types, 
circles to Sab and Sb types, triangles to Sbc's and later types.}
\label{figure:samp}\protect
\end{figure*}

The quantity $r_d^{\circ}$ appearing in Eq. (\ref{eq:c_ri_in}) is the 
face-on disk scalelength measured in the region of the galaxy where the disk 
is actually 
transparent. This in general will be different from $r_d$, the scalelength 
we obtain from our decompositions, since the latter is computed
over the whole disk; we assume that at least for the galaxies which are 
face-on our estimate of $r_d$ can be considered a reasonable approximation 
for $r_d^{\circ}$. 
We correct $r_d$ for the other galaxies using Eq. (\ref{eq:c_r_in}), 
with a coefficient $\eta = 0.5$. This choice for $\eta$ for our $H$ band 
data will be justified later on.
The plot we obtain is shown in Fig. \ref{figure:samp}; 
here and in the rest of the paper Sa galaxies and earlier
are represented by crosses, Sab and Sb by open circles, Sbc and later
types by filled triangles. No obvious trend of $r_{21}^{\circ}$ 
with inclination appears.
To make our sample as fair as possible, spanning approximately the same 
interval in $r_{21}^{\circ}$ at all inclinations, we exclude 12
more galaxies and restrict the following analysis to the objects whose face-on 
isophotal radius lies between 2.9 and 16.9 kpc (dotted lines in 
Fig.~\ref{figure:samp}). The final sample includes 154 galaxies.

If we consider now the correlation of $r_{21}$ (uncorrected to face-on 
aspect) versus $\log (a/b)$ for this subsample, we can obtain an estimate of 
the {\it average} 
$\delta_{21}$ coefficient; we find $\delta_{21} = 0.85\pm 0.16$.
From Eqs. (\ref{eq:c_ri_in}) and (\ref{eq:c_riso_in}) we see that for 
transparent disks, this quantity is $2.306\, r_d^{\circ}/r_{21}^{\circ}$, 
implying $r_{21}^{\circ}= 2.7\, r_d^{\circ}$. 
On the other hand, from the observed ratios $r_d/r_{21}$ in the face-on case 
we can obtain 
an independent estimate for $\delta_{21}$: the intercept at zero inclination
of the best fit to $r_d/r_{21}$ versus $\log(a/b)$ yields again 0.85 
($\pm 0.03$). 
Since this latter value depends also on the disk scalelengths derived from 
the image decompositions, we deduce that for face-on galaxies our estimate 
of $r_d$ is not significantly different from the value expected in the 
case of transparency, as we have previously assumed.

For 151 galaxies of the final sample R.G. and M.H. provided 
21 cm line widths, from their private database (``AGC'')
Further evidence that our sample is not heavily affected by selection 
biases is supported by the fact that we do not observe any trend with 
$\log (a/b)$ of the line widths, corrected for inclination.

\section{Observed correlations with inclination \protect\label{sec:corr}}

When looking at the correlation between an observed parameter and the 
inclination, ideally one should analyze separately galaxies of different 
morphological type and absolute luminosity. Objects belonging to 
different classes, in fact, can be characterized by different extinction
properties and therefore by different slopes in the correlation 
considered (G95, Tully et al \cite{tully}).
Moreover, the observed parameter itself (for example the central disk 
brightness) is likely to depend both on the galaxy luminosity and the 
morphological type (G95; de Jong \cite{dejong}; Morton \& Haynes \cite{rh}),
and not accounting for these trends would introduce additional scatter in the 
correlation with $\log (a/b)$.
In any correlation with inclination considered for our data, we do not observe 
obviously different {\it slopes} between galaxies belonging to different 
classes (we will return to 
this point in Sect. \ref{sect:holm}). Therefore we assume a common slope for 
each correlation with $\log (a/b)$. On the other hand, as described in the 
next section, we account for 
possible trends with luminosity and morphological type of the observed 
parameters, i.e. we allow the {\it offsets} of the correlations 
with inclination to be different for objects of different classes. 
Such trends are rather evident, at least in some cases:
more luminous galaxies are characterized by larger $r_d$'s; Sa's 
are more compact on average (smaller $r_d$ and brighter $\mu(0)$) than
later type galaxies of the same luminosity, and so on.
The $H$-band luminosity, in turn, appears to depend little on inclination,
as we will show in Sect. \ref{sect:holm}, and in the next two sections we 
will consider the observed luminosities, uncorrected to face-on aspect.  

\subsection{Disk scalelength \protect\label{sec:c_rad}}

\begin{figure*}
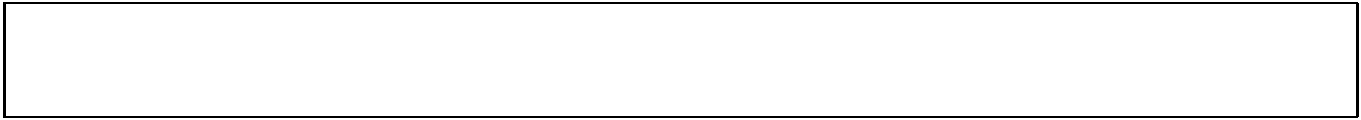

\picplace{1.5cm}
\caption[]{{\bf a}: Disk scalelength plotted versus absolute luminosity,
after rescaling $\log (r_d)$ for each morphology class; the straight line is 
the best fit to the data. 
{\bf b}: residuals from the 
best fit in panel a plotted versus $\log (a/b)$; the average scalelength is
added to each residual in order to preserve the same $y$-axis scale. The best 
fit to $r_d$ versus $\log (a/b)$ is also shown. {\bf c} and {\bf d}: the 
same correlations for the central disk surface brightness.}
\label{figure:rd}\protect
\end{figure*}

Figure \ref{figure:rd}a shows the correlation between $H$-band disk 
scalelengths (on a logarithmic scale) and absolute disk luminosity. 
In order to account at the same time for the dependence of $r_d$ both on 
luminosity and on morphological type, the scalelengths are 
plotted after scaling each morphological type by an additive term, so that 
for every type, the average $M_H$ and the average rescaled $\log (r_d)$ 
lie on the best linear fit to the whole (rescaled) sample. 
The plot is obtained iteratively: first we perform a 
linear regression to the observed  values, after rebinning the points in 
groups of 12; then, for each morphological type, we scale the $r_d$'s by an 
additive term so that the average scalelength for that class lies on 
the linear regression. 
The best fit is then recomputed and a new rescaling accomplished; 
the process is repeated until convergence.
The {\it residuals} with respect to the final best fit are plotted 
versus $\log (a/b)$ in Fig. \ref{figure:rd}b. The average scalelength is 
added to each residual, in order to preserve the same $y$-axis scale;
the best fit of Eq. \ref{eq:c_r_in} to the plot is also shown, yielding a 
slope $\eta = 0.41 \pm 0.17$.
We remind that a correlation of $r_d$ with inclination is to be found
only if some extinction is present, at least in the inner regions 
of the disk.

When all the galaxies are fitted, instead of only those in the selected 
``fair sample'',
we find a slope of 0.5. This is the value we used in Sect. \ref{sec:sel} 
to correct the isophotal radii to their face-on value. 
We find that for any value of $\eta$ from 0.0 (no correlation of 
$r_d$ with inclination) to 0.6 (the value derived by G94 in the $I$ band)
the plot of $r_{21}^{\circ}$ versus $\log (a/b)$ 
(Fig.~\ref{figure:samp}) and the 
subsequent selection of a ``fair sample'' do not change much.
As a consequence, if we plot the disk scalelengths versus inclination for a 
sample selected assuming $\eta = 0$, the value for $\eta$ that we derive a 
posteriori from the plot is not zero but 0.56, implying that an increase of 
$r_d$ with inclination is the only self-consistent possibility.

\subsection{Disk surface brightness}

In principle, the behaviour of the central disk surface brightness $\mu(0)$ is 
rather sensitive to the amount of extinction. In practice, some authors 
(e.g. G94; Byun et al. \cite{byun}) have questioned the actual 
reliability of the observed trends involving $\mu(0)$ for several reasons: 
the large intrinsic 
scatter, difficulties arising in the evaluation of $\mu(0)$ due to the 
presence of the bulge, a shallow dependence of $\mu (0)$ on inclination
unless optical depth is very low. 
In our case, the use of a 2d decomposition of the brightness distribution
and the fact that in the NIR extinction is actually lower than at
optical bands allow us to attribute more significance to this 
test.

Indeed, we do find a correlation between $\mu(0)$ and $\log (a/b)$,
after accounting for possible trends of the disk central brightness
with respect to luminosity and morphological type in the way explained in 
the previous section. The plots are shown in Fig. \ref{figure:rd}c and d. 
The best fit to the data in panel d yields 
$C=0.60 \pm 0.14$, a rather high increase with inclination which 
tends to exclude high disk opacities.
This value is consistent with two published estimates of the coefficient,
both derived using average disk surface brightnesses from aperture 
photometry. In particular Peletier \& Willner 
(\cite{pw}) confined $C_H$ between 0.6 and 0.8, and deduced a central face-on 
optical depth of about 0.2 using a DDP sandwich model. Boselli \& Gavazzi 
(\cite{boselli}) found $C_H  = 0.65$, independent of morphological type.
The face-on central surface brightness turns out to be 17.6 $H$-mag 
arcsec$^{-2}$ with a dispersion of 0.7.

\subsection{The Holmberg test \protect\label{sect:holm}}

Let us consider now the so-called Holmberg test (Holmberg \cite{holm1},
\cite{holm2}), i.e. the 
correlation between inclination and the average surface brightness within 
the circular aperture defined by a certain isophotal radius.
An observed trend of this quantity with inclination could be
interpreted as the signature of significant opacity in the disks.
However, as stated by G94, a trend is expected
also in the case of transparent galaxies because the isophotal 
radius is an increasing function of inclination (except in the case of
{\it completely} opaque disks), so that the surface brightness
tends to decrease at high inclinations because it is estimated using a larger 
aperture.

\begin{figure*}
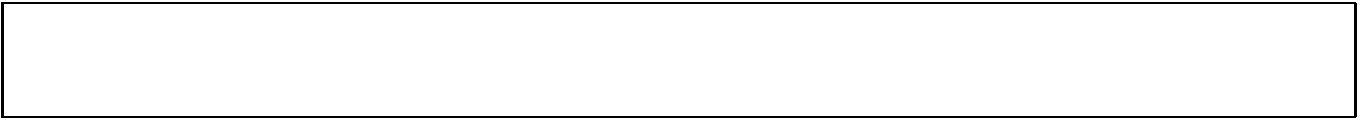

\picplace{1.5cm}
\caption[]{Top panel: Average surface brightness within $r_{21}$ versus
$\log (a/b)$. 
The three arrows represent the trend expected for 
different opacities: the solid arrow is for completely opaque galaxies,
the dotted arrow for completely transparent ones, the dashed arrow
is derived from the observed slopes of $r_d$ and $\mu(0)$ versus 
inclination.
Bottom panel: average surface brightness within $r_{21}^{\circ}$ versus
$\log (a/b)$, with the best fit to the data points.}
\label{figure:holm}\protect
\end{figure*}

Figure \ref{figure:holm} (top panel) displays the correlation found for our 
data; in this case, each morphology class was scaled by an additive
term such as to have a common average on the $y$ axis. Also shown are
the expected slopes in the case of a completely transparent disk (dotted 
arrow), a completely opaque one (solid arrow), and
the behaviour expected when isophotal radius and disk scalelengths vary 
with inclination according to the relations described in the previous 
sections. 
The three slopes are computed for a pure exponential and assuming 
$r_{21}^{\circ}/r_d^{\circ}=2.7$. 
We note that the coefficients derived from our data yield a trend 
which is nearly the same, in terms of average surface brightness, to 
the one expected for completely transparent galaxies.
Actually, as 
already stated by G94, this kind of correlation is not particularly 
sensitive to the amount of internal extinction, and the scatter in the 
plot is too large to distinguish reliably between the different cases.

Following G94, we also consider a modified version
of the Holmberg test in which the surface brightness is calculated 
within the {\it face-on} value of the isophotal radius. In particular, 
in the bottom panel of Fig. \ref{figure:holm}, we plot the disk 
magnitude plus $5 \log r_{21}^{\circ}$ versus the inclination, and the
best fit to the data. Since $r_{21}^{\circ}$ -- computed using Eq.
\ref{eq:c_riso_in} with $\delta_{21} = 0.85$ --
is by definition independent of inclination, the slope of this plot
is the coefficient $\gamma$ in Eq. (\ref{eq:c_m_in});
the value of $\gamma$ corresponds roughly to the correction 
for the total magnitude from edge-on to face-on aspect.
We find little correlation between the two quantities ($\gamma = 0.16 \pm
0.22$), which, at first sight, is the signature of nearly complete transparency.
Actually, this small correlation is consistent with the other trends 
observed: the brightening of $\mu(0)$ with inclination in our galaxies is 
smaller than in the transparent case, yet the increase of $r_d$ -- which 
does not occur if disks are optically thin -- contributes to raise the value 
of the total luminosity within $r_{21}$, making the average surface brightness
behave very much like in the case of complete transparency.
Quantitatively, if we parametrize the correlation between $r_d$ and 
$(a/b)$ as a 
power law, instead of using Eq. \ref{eq:c_r_in}, in particular if we write
\begin{equation}
\protect\label{eq:c_r_in2}
r_d = r_d^o \left( \frac{a}{b} \right)^{\alpha} \;\; ,
\end{equation}
it can be easily demonstrated that $\gamma = 1 - C - 2\alpha$.
For our data we find $\alpha = 0.2 \pm 0.05$, yielding $\gamma = 0 \pm 0.15$,
consistent with the value obtained from Fig. \ref{figure:holm}.

In recent work, Tully et al. (\cite{tully}) derived estimates for the 
extinction correction to face-on both in the optical and in the NIR ($K'$ 
band), as a function of galaxy luminosity.
Assuming $H-K' \sim 0.2$ mag, and since the median absolute
luminosity for the galaxies in our sample is $-23.7$ $H-$mag, we derive from 
their Eq. (6) $\gamma_{K'} = 0.26$, a value which is consistent
with ours, if we assume the difference between $\gamma_{K'}$ 
and $\gamma_{H}$ to be negligible.
A dependence of the amount of internal extinction on the galaxy luminosity
is also found by G95. For our galaxies we do not observe any trend of this 
kind, but we note that our sample
covers a limited range in absolute magnitude (about three magnitudes, and 
even less when considering the galaxies at high inclination). Moreover, 
for galaxies as luminous as ours (we estimate our $I$-band absolute magnitudes 
to be brighter than $-20$), the G95 data suggest only a rather shallow 
dependence of extinction on luminosity (see e.g. their Fig.~7).

\subsection{Bulge parameters}

We look now for correlations of the $H$-band bulge parameters with 
inclination, in particular of the bulge effective radius $r_e$, effective 
surface
brightness $\mu_e$ and absolute magnitude $M_b$, using the same approach 
outlined for the disk parameters in the previous sections.
We find a small increase of $r_e$ with inclination, and little or no 
correlation for $\mu_e$ and $M_b$.
In particular, if we define three coefficients $\eta_b$, $C_b$ and 
$\gamma_b$ using the same relations adopted for the disk 
(Eqs. \ref{eq:c_r_in}--\ref{eq:c_m_in}), we 
derive from the best fits to the data the following values:
\begin{eqnarray}
\eta_{b}   & = & 0.24 \pm 0.16  \\
C_b        & = & 0.04 \pm 0.16  \\
\gamma_{b} & = & 0.09 \pm 0.55  \;\; .
\end{eqnarray}
The bulge-to-disk ratio as well exhibits no significant dependence on 
inclination.

In this case a straightforward interpretation of the results is more 
difficult than for the disk parameters: in fact the bulge and the 
dust are characterized by completely different spatial distributions, 
respectively a peaked and compact spheroid and an extended thin disk,
so that in general extinction will not be uniformely distributed 
throughout the spheroid. 
In the transparent case, however, we would expect the scalelength to 
be independent of inclination, and the effective surface brightness
to increase with $\log (a/b)$ (though not as much as $\mu (0)$, due to the 
lower intrinsic ellipticity of the bulge). Since this is not the case 
for our sample, we interpret the different observed behaviour, again, as the 
effect of some extinction, that needs to be present at these wavelengths,
at least in the central galaxy regions. 

\subsection{$I-H$ colour profiles \protect\label{sec:colour}}

For 68 of our galaxies, $I$ band brightness profiles are available from the 
ScI sample of Haynes et al. (in preparation);
for these objects, we 
compute $I-H$ colour profiles. To reduce them to a common radial scale, 
we normalize the distance from the center with the value of the disk 
scalelength, measured in the $H$ band and corrected to face-on aspect using 
Eq. (\ref{eq:c_r_in}). 

\begin{figure*}
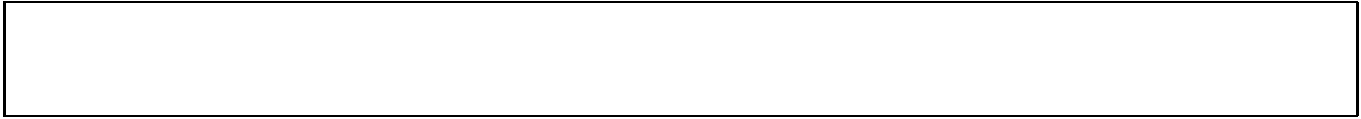

\picplace{1.5cm}
\caption[]{$I-H$ radial colour profiles averaged in three bins of absolute 
magnitude. Radii are normalized with the disk scalelength corrected 
to face-on aspect.}
\label{figure:cpro}\protect
\end{figure*}

A reddening of the observed colours with inclination at a given radius 
and for a fixed stellar population
is expected if (and only if) some dust is present, since
the optical depth along the line of sight increases with inclination, 
and at shorter wavelengths the scattering is more effective in 
removing photons traveling within the plane of the disk. 
Since the dust is concentrated towards the center of the galaxy it 
will also produce a radial colour gradient, making the outer regions of
the galaxy look bluer. However, a contribution to such a gradient 
can also be provided by a variation of metallicity or average stellar 
population with radius. Similarly, higher extinction, metallicity, and 
different stellar content, can be invoked to explain the redder colours of high 
luminosity galaxies (Gavazzi, \cite{gavazzi}; Tully et al. \cite{tully}).  
On the other hand, a correlation of colour with inclination provides direct 
and unambiguous evidence for reddening by dust.
Figure \ref{figure:cpro} shows the average $I-H$ radial profiles for our 
sample in three bins of absolute magnitude. As expected, they become bluer
with increasing radius, and brighter galaxies tend to be redder at any given
radius.

\begin{figure*}
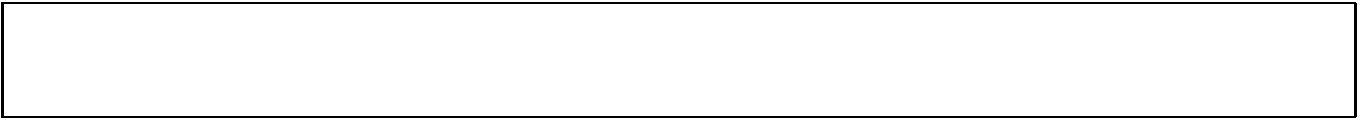

\picplace{1.5cm}
\caption[]{Left hand panels: colour-luminosity relations. In each panel, from 
top to bottom, we plot $I-H$ colours averaged in a different radial bin. 
The best fit to the data are also shown. 
Right hand panels: the residuals from the best fit to the colour-luminosity 
relations in the left hand panels are plotted versus the inclination.
In each panel the average colour for that particular bin is added to each 
residual, to preserve the same $y$-axis scale. The best fits to the data 
are also shown.} 
\label{figure:ac}\protect
\end{figure*}

We inspect for trends of $I-H$ with inclination in three radial bins:
within one disk scalelength, between one and two, and between two and 
three $r_d$'s respectively. 
We exclude the region inside 1/3 of $r_d$, where different seeing between 
$I$ and $H$ and the presence of a bulge can produce extremely high "local" 
colour gradients.
Since the reddening of colour with increasing luminosity would introduce 
scatter in the correlation with inclination, in each radial bin we 
obtain a best fit to the colour-luminosity relation and plot the 
residuals versus the inclination.
We do not observe any difference in colour between early and late-type 
spirals; therefore in this case we didn't rescale the different morphological 
types. 
Figure \ref{figure:ac} (left hand panels) shows the observed colour-luminosity 
relations in the 3 radial 
bins with a best fit for each bin; the slopes are $-0.12 \pm 0.01$,
$-0.11 \pm 0.02$ and $-0.08 \pm 0.03$
respectively; the sign of the slopes is the one expected (redder colour 
with increasing luminosity). In the right hand panels we plot the residuals
of these correlations versus $\log (a/b)$.
In each panel the average colour for that particular bin is added to each
residual, to preserve the same $y$-axis scale; the best fits to the data
are also shown.
As expected we find that more inclined galaxies tend to be redder.
The slopes of the best fits are 
$0.34 \pm 0.04$ mag, $0.56 \pm 0.07$, and $0.6 \pm 0.1$ 
respectively.

If we consider the average disk colours, the slope in the correlation between
colour and inclination amounts to $\gamma_I - \gamma_H$, and we find this 
slope to be 0.46~$\pm$0.14. Again, we can compare our result to the values 
found by Tully et al. (\cite{tully}) and G94. The median $I-H$ for our sample 
is 1.74, yielding for the $I$ band luminosity a value of about $-22$. 
Equation (5) in 
Tully et al. then gives $\gamma_I = 1.05$, which is also the value 
found by G94 for their sample. 
Our data seem to suggest a slightly lower value for $\gamma_I$ (i.e. lower 
extinction), for which we can 
set an upper limit of about 0.9 from the colour-inclination slope and from
the value of $\gamma_H$ derived in the previous section.

\subsection{Color gradients and inclination \protect\label{sec:cgrad}}

\begin{figure*}
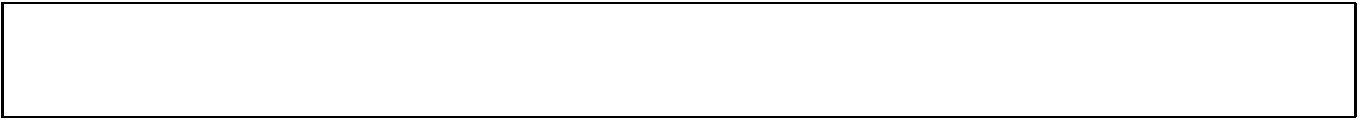

\picplace{1.5cm}
\caption[]{Radial colour gradients versus $\log (a/b)$. The dashed line 
is the best fit to the data assuming a power-law dependence of $r_d$ on the 
axial ratio $a/b$ both in $I$ and $H$.}
\label{figure:cgrad}\protect
\end{figure*}

We also measured colour gradients by fitting to each profile a straight line,
still excluding the points inside 0.3 $r_d^{\circ}$. 
Figure \ref{figure:cgrad} shows the gradients obtained for this subsample 
plotted versus inclination. 
Assuming that an exponential disk fits well both the $I$ and $H$ brightness
distributions, it can be easily shown that 
\begin{equation}
\protect\label{eqn:cgrad}
\frac{d\, (I-H)}{d\, \rho} = 1.086\, \left(\frac{r_H}{r_I} - 1\right) 
\end{equation}
where we have denoted with $r_I$ and $r_H$ the disk scalelengths in the 
two bands, and $\rho = r/r_H$ is the radius in units of the $H$ band 
disk scalelength. If we parametrize the dependence of $r_d$ on inclination
with Eq. \ref{eq:c_r_in2}, we obtain
\begin{equation}
\protect\label{eqn:cgrad2}
\frac{d\, (I-H)}{d\, \rho} = 1.086\, \left(\frac{r_H^o}{r_I^o}
\left( \frac{a}{b} \right)^{\alpha_H - \alpha_I} - 1\right) \;\; .
\end{equation}
The fact that a negative gradient is almost always detected implies
that on average $r_H$ is smaller than $r_I$; moreover, this difference 
is about the same at all inclinations -- amounting to 10 $\sim$ 20\%
-- which is what we expect if the disk scalelengths correlate in a similar 
way with inclination in both bands, i.e. $\alpha_H \simeq \alpha_I$.
If any correlation is to be found in Fig.~\ref{figure:cgrad}, it is in the 
sense of smaller
gradients for high inclinations, implying $\alpha_H > \alpha_I$ from Eq. 
(\ref{eqn:cgrad2}).
Looking at Fig.~7 in G95 this seems actually to be the case for 
galaxies brighter than $-20$ $I$-mag, for which a value of $\alpha_I \simeq 
0.1$ is appropriate, whereas we found $\alpha_H = 0.2$. 
Note that from the same figure it turns out that the dependence on 
inclination of disk scalelengths is steeper ($\alpha$ is higher) for 
galaxies of low luminosity, which the authors claim to be {\it less}
extincted. The fact that $\alpha$ is higher also in the $H$ band
where extinction is certainly lower is therefore a confirmation of 
G95's result. 
A best fit to the data points in Fig. \ref{figure:cgrad} (dashed line) 
yields $\alpha_H - \alpha_I = 0.11 \pm 0.04$, and a rather accurate
estimate for the ratio $r_H^o/r_I^o = 0.81 \pm 0.03$.

\section{Disk opacity in the NIR}

Some qualitative statements can be made on the basis of the correlations
shown above. It appears that some of the parameters which describe 
the brightness distribution of a galaxy are affected
by the presence of interstellar dust also in the NIR. At these wavelengths,
however, the effect of extinction on most observable quantities is quite
small. The best evidence for the presence of extinction is in the correlation
between disk scalelength and inclination -- which 
turns out to be similar to the one existing in the $I$ band
(Sect. \ref{sec:cgrad} and Sect. \ref{sec:c_rad}) -- and in that between 
$I-H$ colour and inclination.  On the other hand, the behaviour of 
the disk central brightness is not very different from what we expect in 
the case of transparency; such a correlation, as pointed out by Byun at al, 
\cite{byun}, is to be seen only when opacity is low. Moreover, we find 
little variation of the total luminosity with inclination, and the 
correction from edge-on to face-on for the total magnitude is about
0.15 mag on average confirming that extinction in the NIR , though detectable, 
is not conspicuous even at high inclinations.

\subsection{Simulating absorption by dust}

In order to better quantify our estimate of optical depth in the disks,
we simulate a set of brightness distributions affected by absorption 
by dust for a range of inclinations and different values of the central 
optical depth. We adopt a model galaxy in which dust and stars
are distributed exponentially both in the radial and ``vertical'' 
direction, like in the triplex model by DDP, but
we assume the dust scalelength to be larger than $r_d$,
following Peletier et al. (\cite{pelet}), the results of two recent
works by Xilouris et al. (\cite{xiloa}, \cite{xilob}), and 
of Nelson et al. \cite{nelson}. 
In particular our dust disks are 1.5 times larger than the stellar ones.
The models are major-axis brightness profiles computed according to the 
approximate analytical solution for radiative transfer in a dusty disk
described in DDP. We consider pure disks (no bulge component)
with dust-to-stars scaleheight ratio $\zeta$ = 0.25, 0.5, or 0.75, and 
$\tau_H(0) \leq 2$. 
Scattering effects are not included.
The profiles are fitted with exponentials excluding the inner part, 
within 0.3 disk scalelengths, i.e. the region in our galaxies where the 
bulge is usually the brighter component. To make the fitting
process as similar as possible to the one we performed on real galaxies,
a proper weighting is given to the various points of the profile
in order to simulate the typical errors in our data.
Since our sample is not uniformly distributed in $\log (a/b)$, we 
define a set of inclinations for the models by rebinning the values for 
the real data in groups of 12 galaxies each. The highest inclination bin 
is centered at $80^{\degr}$, where the numerical approximations in the model 
are still reasonably accurate (see DDP and G94).

We also compute a set of models to simulate the bulge brightness profiles
in the region where they are usually brighter than the disk. For our galaxies
the bulge is in general well represented by an exponential, and it dominates
the brightness distribution at any inclination within approximately two
effective radii (i.e. about three bulge scalelengths $r_b$, or 
0.3 disk scalelengths). The median intrinsic ellipticity is 0.4.
A satisfactory modeling of the bulge-dominated region of a galaxy should 
in principle be carried out using two dimensional brightness distributions,
to analyze the information provided by the asymmetries, due to extinction,
between the two halves of the galaxy image. It should also account for the 
presence of the underlying disk distribution.
In our case we will consider our simple model of the major-axis surface 
brightness profile only as a rough check for the observed 
trends of the bulge parameters with inclination. In this perspective we 
adopt the same type of model we used for the disk (the product of two 
exponentials, along the radial and vertical direction respectively),
though a density distribution characterized by ellipsoidal isophotal 
surfaces would be more appropriate to represent the spheroidal
shape of the bulge.
For this set of models we choose a dust scalelength 13 times larger than $r_b$;
using the typical intrinsic bulge ellipticity (0.4) and the bulge-to-disk 
scalelength ratio ($\sim 0.1$) of our sample galaxies, we also transform the 
dust-to-stars scaleheights adopted for the disk, namely 0.25, 0.5, and 0.75,
into 0.35, 0.7, and 1 respectively.

\subsection{Disk and bulge parameters}

\begin{figure*}
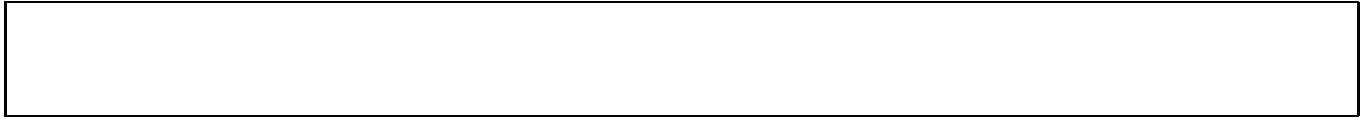

\picplace{1.5cm}
\caption[]{From top to bottom: $H$-band $\eta$, $C$, and $\gamma$ coefficients
versus central optical depth. Dashed lines represent 
simulations with dust-to-stars scaleheight ratio $\zeta = 0.25$; for thin solid 
lines $\zeta = 0.5$, and for dotted lines $\zeta = 0.75$. The shaded areas 
are the ranges we derive from the observed disk parameters at 1 $\sigma$ 
confidence level.}
\label{figure:sim}\protect
\end{figure*}

Figure \ref{figure:sim} shows the expected slopes of the correlations of 
disk scalelength, central brightness, and total luminosity with inclination 
(coefficients $\eta$, $C$, and $\gamma$). 
The coefficients are plotted as a 
function of central optical depth. In each panel the dashed line, the solid 
line, and the dotted line correspond to the simulations 
with $\zeta =$ 0.25, 0.5 and 0.75 respectively.
The shaded areas are the estimates of the coefficients derived from
our sample at 1 $\sigma$ confidence level.

It turns out that all the observed coefficients are consistent with the 
models in a range of central optical depths. In particular,
from panel b, we can set a lower limit for $\tau_H(0)$ around 0.3 
regardless of the thickness of the dust layer, since the three coefficients 
considered exhibit little dependence on $\zeta$ for $\tau_H(0) \la 0.5$. 
From panel c, on the other hand, we derive an upper limit 
to the optical depth around 0.5 again roughly independent of $\zeta$.
We note that our lower limit implies a larger $\tau_H(0)$ than the values 
around 0.1 suggested by the results in Xilouris et al. (\cite{xiloa},
\cite{xilob}). 
Such values, however, are derived for two particular edge-on galaxies 
and not from a statistical analysis, and we do not consider the discrepancy 
particularly worrisome.
Adopting the extinction curve reported by Gordon et al. \cite{gordon} for
the Milky Way, our limits for $\tau_H(0)$ imply a central optical depth 
between 0.9 and 1.4 in the $I$ band, between 1.8 and 3 in the $V$ band,
and between 2.4 and 3.9 in the $B$ band. In agreement with most of the
recent works on the subject, therefore, our data suggest that spiral disks 
are characterized by a moderate opacity also at optical wavelengths.
In particular the presence of a significant extinction in the outer disk 
regions can be safely ruled out, except in the case of extremely inclined 
galaxies.
On the other hand, the extinction corrections derived from
our data cannot be easily extrapolated to the $B$ or $V$ band on the basis
of the extincton curve alone, since they also depend on the exact geometry of 
dust and stars, and on the wavelength-dependent effect of scattering
(see for example Byun et al. \cite{byun}).

In our model we assumed a dust disk larger by 50\% than the stellar 
one. If we adopt equal scalelengths for the two components we find that 
the predicted slopes are consistent with our data for slightly higher values 
of $\tau_H(0)$, namely between 0.4 and 0.8.
If we consider instead an even more extended dust layer, in particular 
if the dust scalelength in twice as large as $r_d$, again we find agreement 
between models and data for $\tau_H(0)$ around 0.3. 

\begin{figure*}
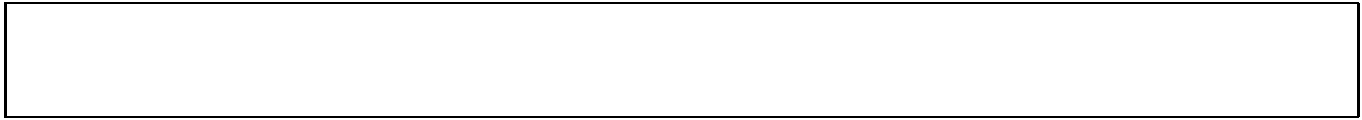

\picplace{1.5cm}
\caption[]{From top to bottom: $H$-band $\eta_b$, $C_b$, and $\gamma_b$ 
coefficients versus central optical depth. Dashed lines represent 
simulations with dust-to-stars scaleheight ratio $\zeta = 0.35$; for thin 
solid lines $\zeta = 0.7$, and for dotted lines $\zeta = 1$. The shaded areas 
are the ranges we derive from the observed bulge parameters at 1 $\sigma$ 
confidence level.}
\label{figure:bsim}\protect
\end{figure*}

Figure \ref{figure:bsim} is the analogous of Fig. \ref{figure:sim} for the 
bulge parameters. Comparing the model predictions in the two figures the 
largest difference is to be found in the behaviour of the coefficient 
$C$, which for the disk is equal to 1 in the transparent case and then 
decreases tending to zero as $\tau(0)$ increases. For the bulge the 
apparent axial ratio is always $\geq b/a$, so that in the 
transparent case $C_b < 1$; moreover, since the dust distribution is much 
more extended than the bulge in the radial direction, for sufficiently high 
optical depths the surface brightness 
along the major axis becomes {\it fainter} with increasing inclination, 
yielding $C_b < 0$.

As explained in the previous section these simulations
are to be considered less reliable than the ones carried out for the disks,
and actually there are no values of the central optical depth for which we have
agreement between models and data for all the three correlations considered,
at 1 $\sigma$ confidence level.
We note however that at 2 $\sigma$ any $\tau_H(0) \la 0.5$ is consistent 
with the data; as was the case for Fig. \ref{figure:sim} this is true 
regardless of the dust-to-stars scaleheight ratio.
This range of optical depths, in turn, is consistent with the one derived 
from the disk correlations, so that we can consider the results from 
these tests as a confirmation -- though at a lower confidence level -- 
of our previous findings.
 
\subsection{$I-H$ colours}

\begin{figure*}
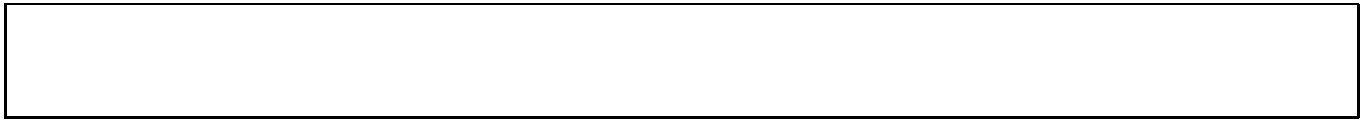

\picplace{1.5cm}
\caption[]{Slope of the correlation of $I-H$ colour versus inclination,
plotted versus central $H$-band optical depth, in three radial bins. 
Dashed lines represent simulations with dust-to-stars scaleheight ratio 
$\zeta = 0.25$; for thin solid lines $\zeta = 0.5$, and for dotted lines 
$\zeta = 0.75$. The shaded areas are the ranges we derive from our data at 
1 $\sigma$ confidence level.}
\label{figure:csim}\protect
\end{figure*}

Figure \ref{figure:csim} shows the dependence on $\tau_H(0)$ of the 
simulated slopes of $I-H$ colour versus $\log (a/b)$ in the three radial bins
considered for the real data. The values we derive from our galaxies are 
represented by the shaded areas. In the simulations $\tau_I$ is assumed to 
be 2.87 times the value in the $H$ band, according to the value reported by
Gordon et al. (\cite{gordon}) for the Milky Way. 
From this plot, we infer that the slopes derived from our data are consistent 
with the limits previously set on $\tau_H(0)$ (0.3 $\sim$ 0.5) only if 
$\zeta < 0.5$. For thicker dust layers the dependence of the colour on 
inclination tends to be too steep within $r_d$, unless we increase the 
central optical depth above 1.
This result holds regardless of the radial extension of the dust distribution 
with respect to the stellar scalelength.
A thick dust distribution appears unlikely also because of the 
little difference observed in the dependence of $r_d$ on inclination 
between the $H$ and $I$ band, i.e. for two optical depths
differing by about a factor of three (see Sect. \ref{sec:cgrad}). 
From Fig.~\ref{figure:sim}a, 
we see that $\eta$ is constant or decreasing with increasing $\tau$ only 
for {\it low} values of $\zeta$ (we neglect differences in $\zeta$ between $H$
and $I$: e.g., Xilouris et al. 1997 find this ratio to be fairly constant
from $K$ to $B$ for NGC~2048).

%
\subsection{Is scattering important?}

In our simple model for an internally-extincted galaxy, we have neglected 
the effect of scattering of radiation by dust; roughly speaking, this means 
that for a given amount of dust (i.e. for a given $\tau(0)$) our model galaxy 
is less luminous than a real galaxy with the same structural parameters. 
How does this affect the conclusions derived from our simulations?
We note that we are not interested on how scattering affects the observed 
quantities, but on how it affects their correlations with inclination.    
A few Monte Carlo simulations of a dusty spiral galaxy kindly provided 
by S. Bianchi allowed us to quantify approximately the differences introduced 
by scattering in the observed $H$-band brightness distribution 
at low and high inclinations. 
The code which performed the simulations, described in Bianchi et al. 
(\cite{bianchi}), produces a set of images
at different inclinations for a given galaxy computing the radiative
transfer without introducing any approximation and including both absorption 
and multiple scattering by dust. 
For every simulated galaxy image the corresponding image including only 
absorption was also computed.
A comparison of the two sets of images (with and without scattering) for a 
galaxy with the average structural properties of our sample and a central
optical depth $\tau_H(0) = 1$ reveals that the contribution by scattering to 
the observed surface brightness is slightly higher at low inclinations and 
increasing toward the center. Yet, the implied corrections to $r_d$ and 
$\mu(0)$ are rather insensitive to inclination when compared to the 
observed variations of these parameters with $\log (a/b)$. 
We estimate the correction to the central disk brightness due to 
scattering to be roughly 0.06 $H$-mag arcsec$^{-2}$ larger in the face-on 
case than in the case of a highly inclined galaxy ($\sim 80^{\degr}$), 
and the decrease in $r_d$ to be around 2\% larger in the face-on case. 
Both these corrections are negligible, since the observed correlations
with inclination for these parameters imply a brightening of $\sim$ 1.15 mag
for $\mu(0)$, and an increase of $\sim$ 30\% for $r_d$, between face-on and 
80$^{\degr}$ inclination. 
The difference in the scattering correction to the total magnitude 
is more significant ($\sim 0.04$ mag) given the small variation of this
quantity with $\log (a/b)$, but still rather small.
Moreover, since the central optical depth is likely to be less than 1 in the 
$H$ band, the actual corrections are probably even smaller than these.
The effect of scattering is likely to be more important when considering 
the correlations with inclinations of the {\it bulge} parameters, given the 
little dependence on $\log (a/b)$ we find for all of them, and the highly 
concentrated bulge brightness distribution. On the other hand, in this
case our model is a rather poor approximation anyway, and we considered
its predictions only as a rough confirmation of the results derived 
from the disk parameters.

In conclusion, we expect that the inclusion of scattering in the model 
would not produce significant changes in our estimate of $\tau_H(0)$.

\section{Summary}

In this paper, we have derived NIR ($H$ band) structural parameters for a 
sample of 174 spiral galaxies in the Pisces-Perseus supercluster, using a 
bi-dimensional decomposition of the galaxy images. We have subsequently
analyzed the correlations of such parameters with inclination in order to 
detect possible effects of internal extinction and estimate the relative
corrections to face-on aspect. For 68 galaxies 
of the sample $I-H$ colour profiles are also available, which were used to 
the same purpose. 
The main results of this work are: \\

1) The effect of internal extinction in the $H$ band, though small,
can be detected. In particular the increase of the average disk scalelength 
and of the disk $I-H$ colours with inclination can be attributed to the 
presence of the dust.\\

2) We find little dependence of the total disk luminosity on inclination. We
derive an average correction of about 0.15 mag from edge-on 
to face-on aspect in the $H$ band. We find this result to be consistent with 
the correlations we find for $r_d$ and $\mu(0)$ versus inclination.
Little or no correlation with inclination is found for the bulge 
parameters.\\

3) Comparing our results with the predictions of a simple model for extincted
galaxies, we deduce that the average central optical depth in the $H$ band is 
between 0.3 and 0.5 if the dust scalelength is 50\% larger than $r_d$. The 
simulations also suggest that the ratio of the dust-to-stars scaleheights is 
on average lower than 0.5. The effect of scattering on the various correlations 
considered is found to be negligible at these wavelengths. 

\begin{acknowledgements}
We would like to thank S. Bianchi for having provided Monte Carlo 
simulations of dusty spiral galaxies; the referee, P.C. van der Kruit; and C. 
Giovanardi and L. Hunt for useful comments and suggestions. 
This research was partially funded by ASI Grant ARS--96--66.
Partial support during residency of G.M. at Cornell University was obtained via 
the NSF grant AST96-17069 to R. Giovanelli.
\end{acknowledgements}


\begin{thebibliography}{}

\bibitem[1995]{apb}
{Andredakis Y.C., Peletier R.F., Balcells M., 1995, MNRAS 275, 874}


\bibitem[1996]{bianchi}
{Bianchi S., Ferrara A., Giovanardi C., 1996, ApJ 465, 127}

\bibitem[1994]{boselli}
{Boselli A., Gavazzi G., 1994, A\&A 283, 12}

\bibitem[1991]{bhf}
{Burstein D., Haynes M.P., Faber S.M., 1991, Nat 353, 515}

\bibitem[1995]{bf}
{Byun Y.I., Freeman K.C. 1995, ApJ 448, 563}

\bibitem[1994]{byun}
{Byun Y.I., Freeman K.C., Kylafis N.D., 1994, ApJ 432, 114}

\bibitem[1989]{cardelli}
{ Cardelli A.J., Clayton G.C.,  Mathis J.S., 1989, ApJ 345, 245}

\bibitem[1996]{corradi}
{ Corradi R.L.M., Beckman J.E., Simonneau E., 1996, MNRAS 282, 1005}


\bibitem[1990]{davies}
{Davies J.I., 1990, MNRAS 244, 8}

\bibitem[1993]{dpbd}
{Davies J.I., Phillips S., Boyce P.J., et al., 1993, MNRAS 260, 491}

\bibitem[1996]{dejong}
{de Jong R.S., 1996, A\&A 313, 45}

\bibitem[1991]{rc3}
{de Vaucouleurs G., de Vaucouleurs A., Corwin Jr. H.G., et al.,
1991, Third Reference Catalogue of Bright Galaxies (Springer-Verlag, 
New York).}

\bibitem[1989]{disney}
{Disney M., Davies J., Phillips S., 1989, MNRAS 239, 939}

\bibitem[1993]{gavazzi}
{Gavazzi G., 1993, ApJ 419, 469}

\bibitem[1994]{g94}
{Giovanelli R., Haynes M.P., Salzer J.J., et al., 1994, AJ 107, 2036}

\bibitem[1995]{g95}
{Giovanelli R., Haynes M.P., Salzer J.J., et al., 1995, AJ 110, 1059}

\bibitem[1997]{gordon}
{Gordon K.D., Calzetti D., Witt A.N., 1997, ApJ 487, 625}


\bibitem[1958]{holm1}
{Holmberg E., 1958, Medn. Lunds astr. Obs., Ser. 2, No. 136}

\bibitem[1975]{holm2}
{Holmberg E., 1975, in Stars and Stellars Systems, Vol. IX, A. Sandage,
M. Sandage, J. Kristian eds. (University of Chicago, Chicago).}

\bibitem[1994]{jansen}
{Jansen R.A., Knapen J.H, Beckman J.E., et al., 1994, MNRAS 270, 373}

\bibitem[1992]{knapen}
{Knapen J.H., van der Kruit P.C., 1992, A\&A 248, 57}

\bibitem[1998]{moriondo}
{Moriondo G., Giovanardi C., Hunt L.K., 1998, A\&AS 130, 81}


\bibitem[1994]{rh}
{Morton R.S., Haynes M.P., 1994, ARA\&A, 32, 115}

\bibitem[1998]{nelson}
{Nelson A.E., Zaritsky D., Cutri R.M., 1998, astro-ph/9803161}

\bibitem[1992]{pw}
{Peletier R.F., Willner S.P., 1992, AJ 103, 1761}

\bibitem[1995]{pelet}
{Peletier R.F., Valentijn E.A., Moorwood A.F.M., et al., 1995, 
A\&A 300, L1}

\bibitem[1968]{sersic}
{S\` ersic J.L., 1968, Atlas de Galaxias Australes
(Cordoba: Observatorio Astronomico)}

\bibitem[1998]{tully}
{Tully R.B., Pierce M.J., Huang J., et al., 1998, astro-ph/9802247}

\bibitem[1997]{xiloa}
{Xilouris E.M., Kylafis N.D., Papamastorakis J., et al., 1997, A\&A 325, 135}

\bibitem[1998]{xilob}
{Xilouris E.M., Alton P.B., Davies J.I., et al., 1998, A\&A 331, 894}

\bibitem[1990]{vale}
{Valentijn E.A. 1990, Nat 346, 153} 

\end{thebibliography}
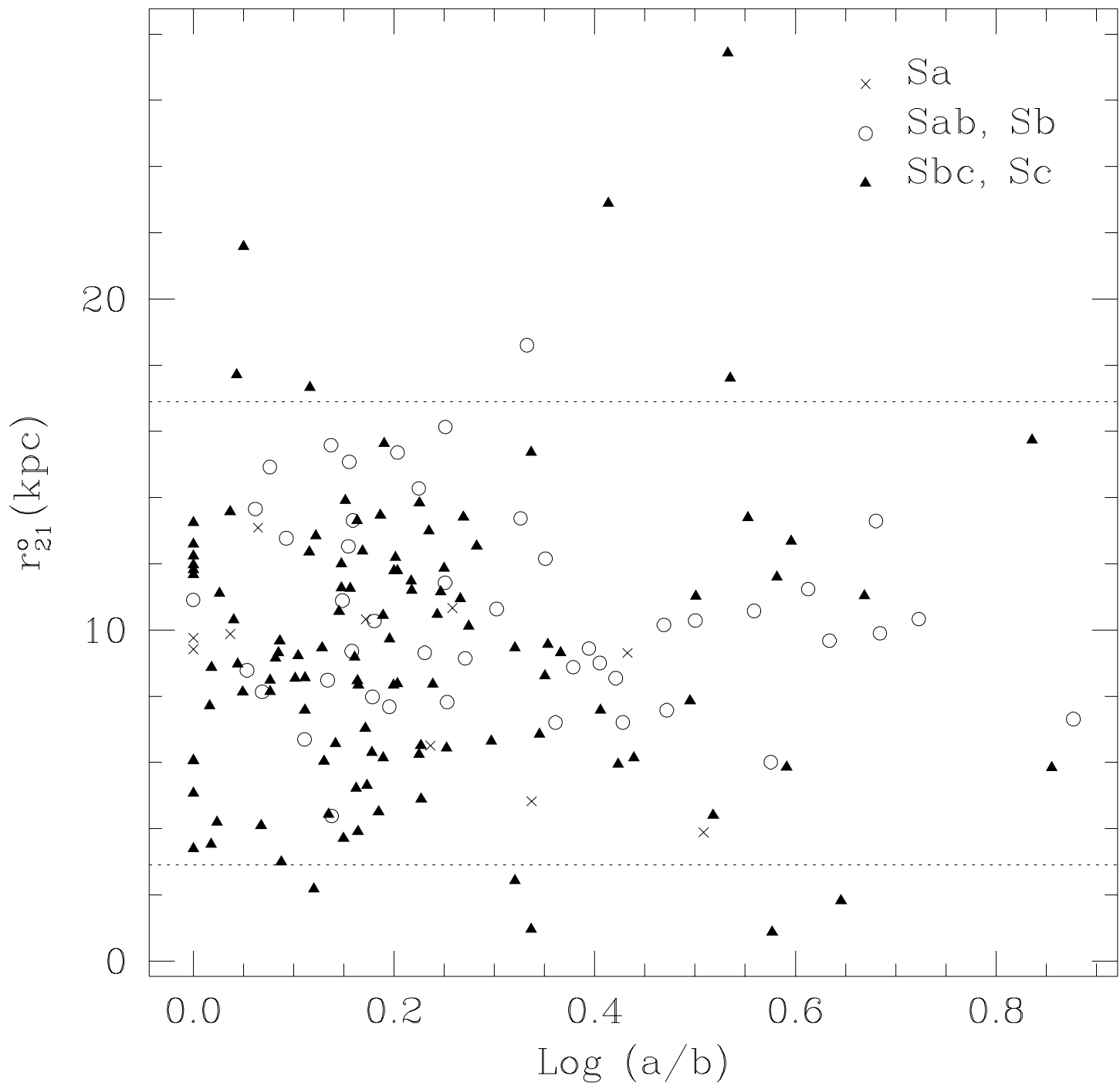
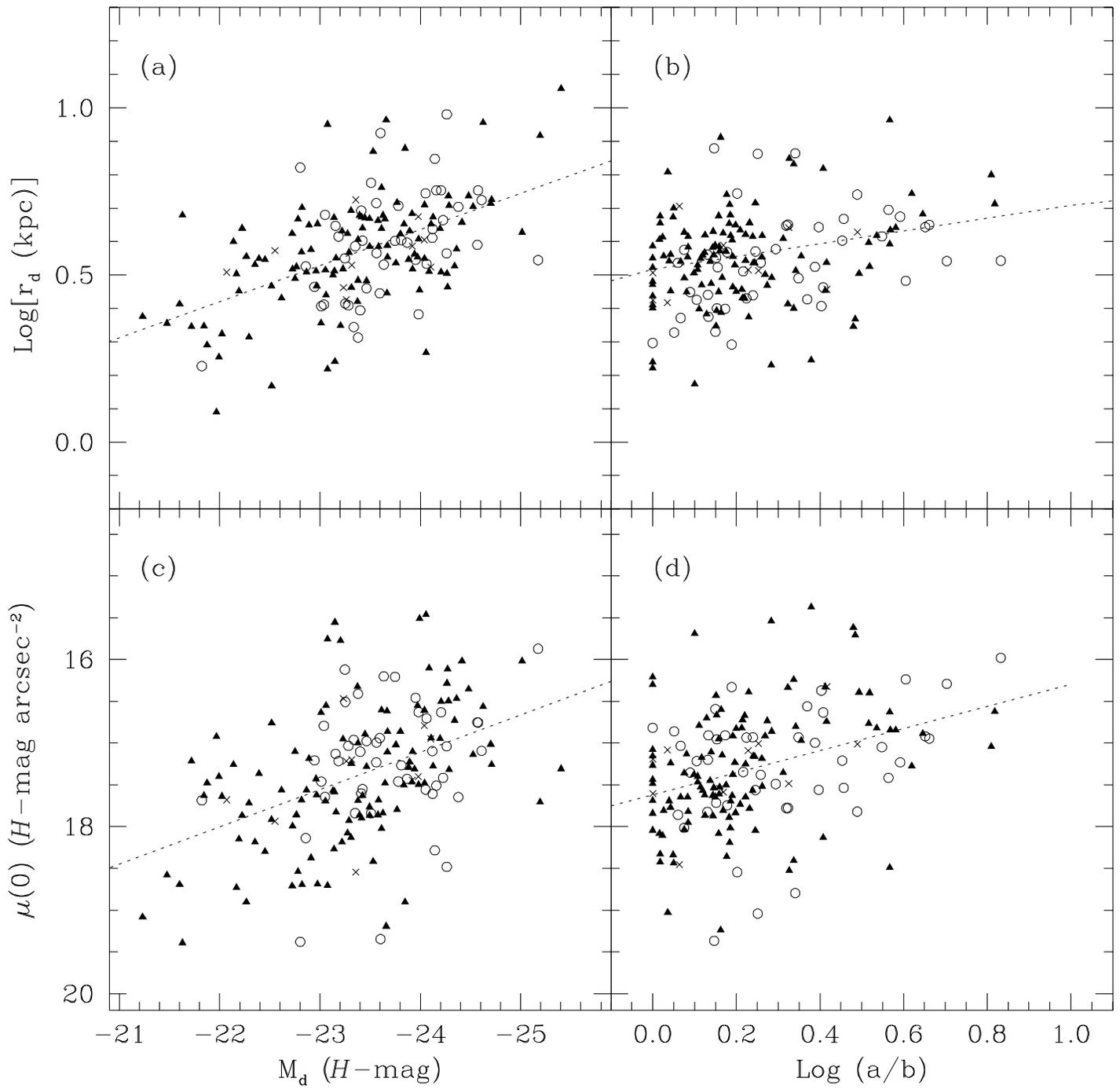
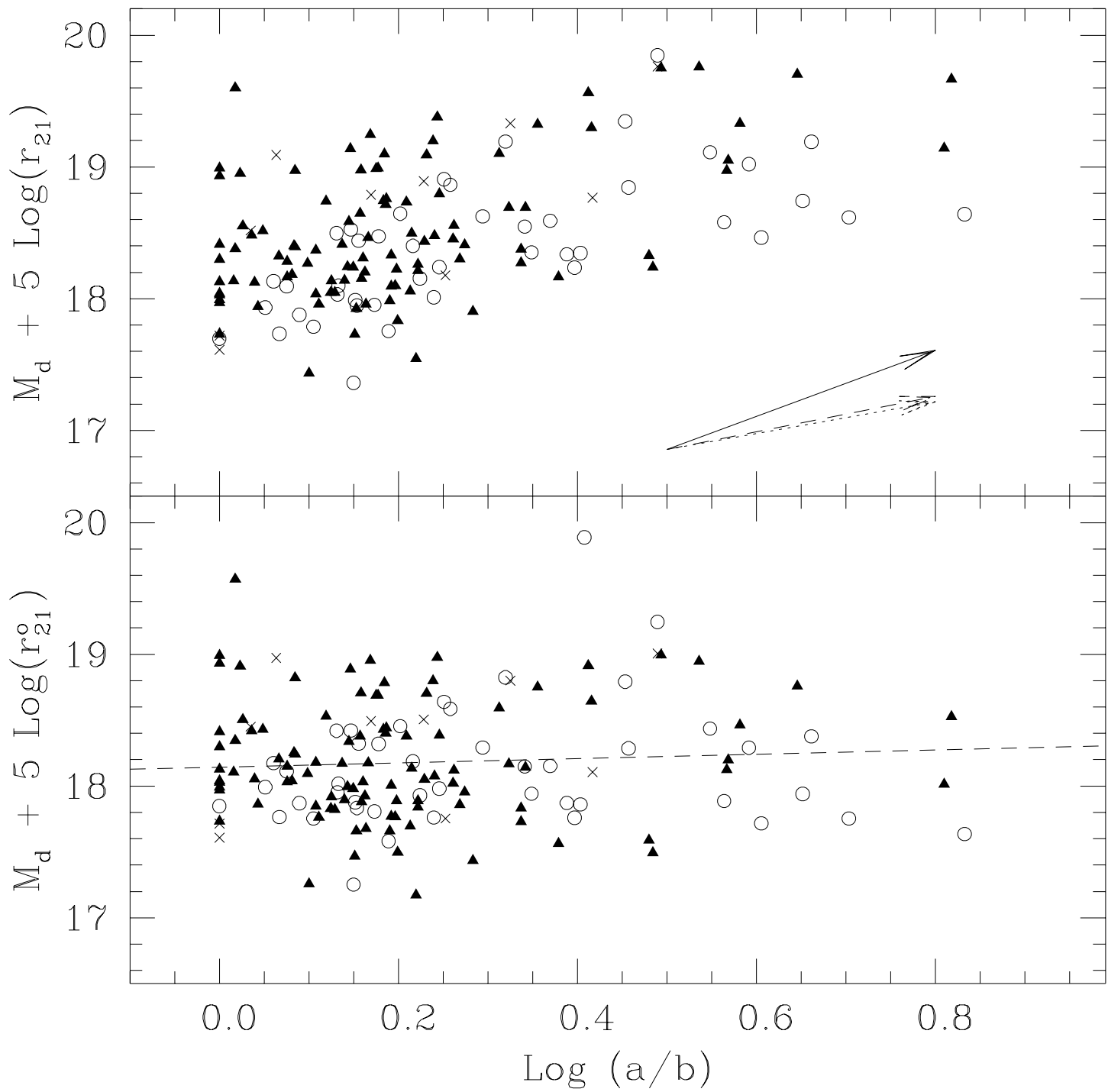
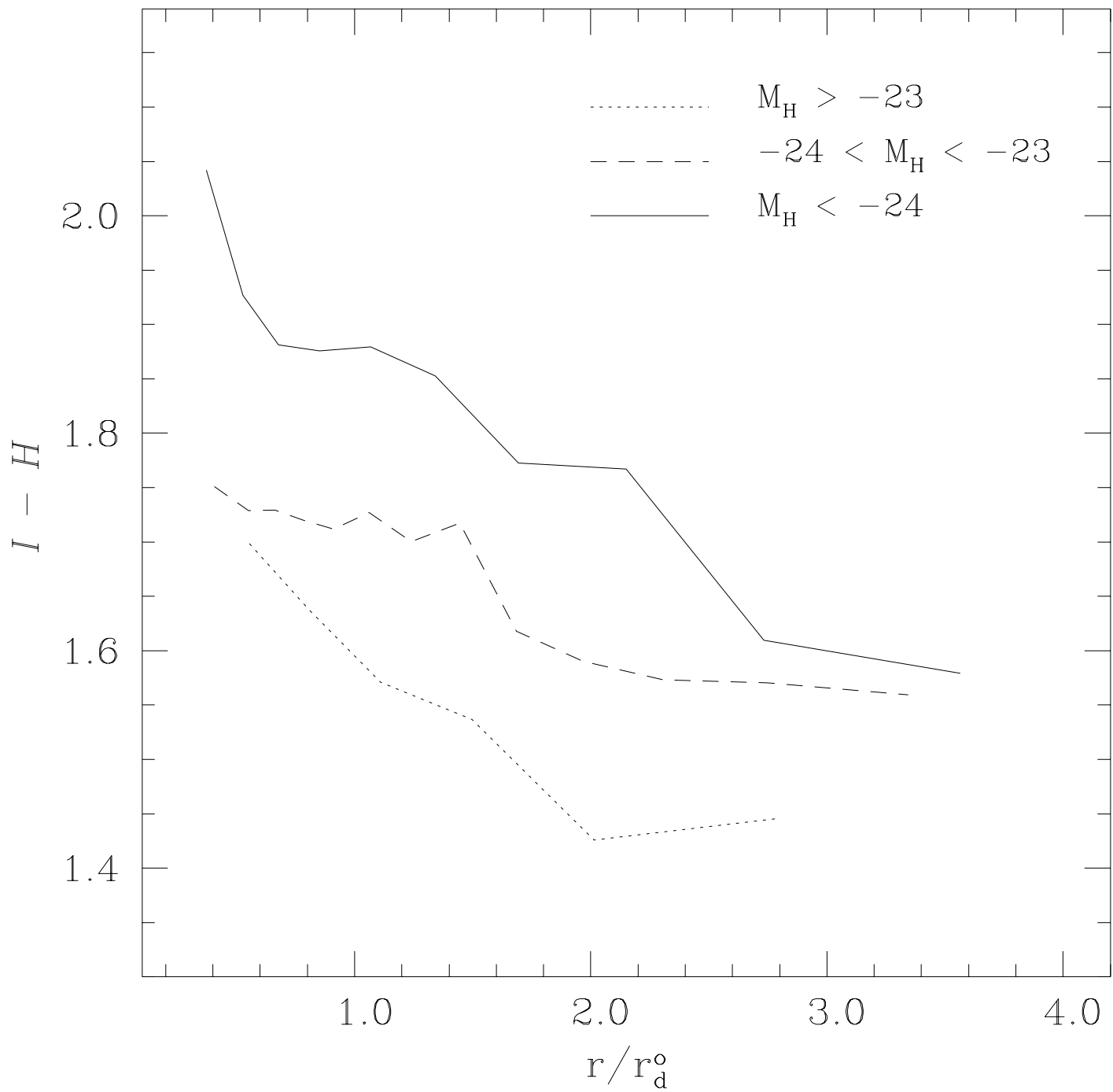
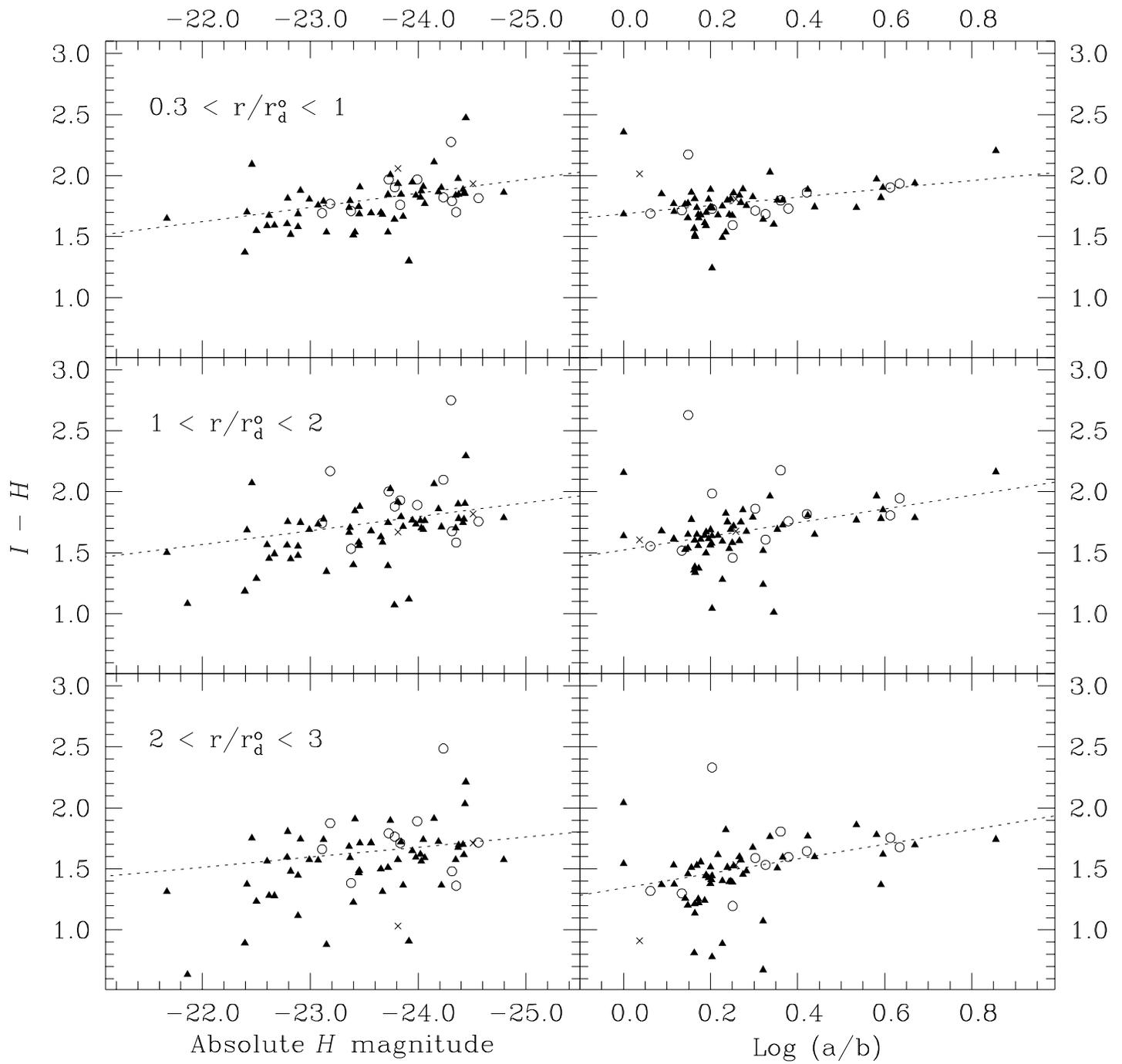
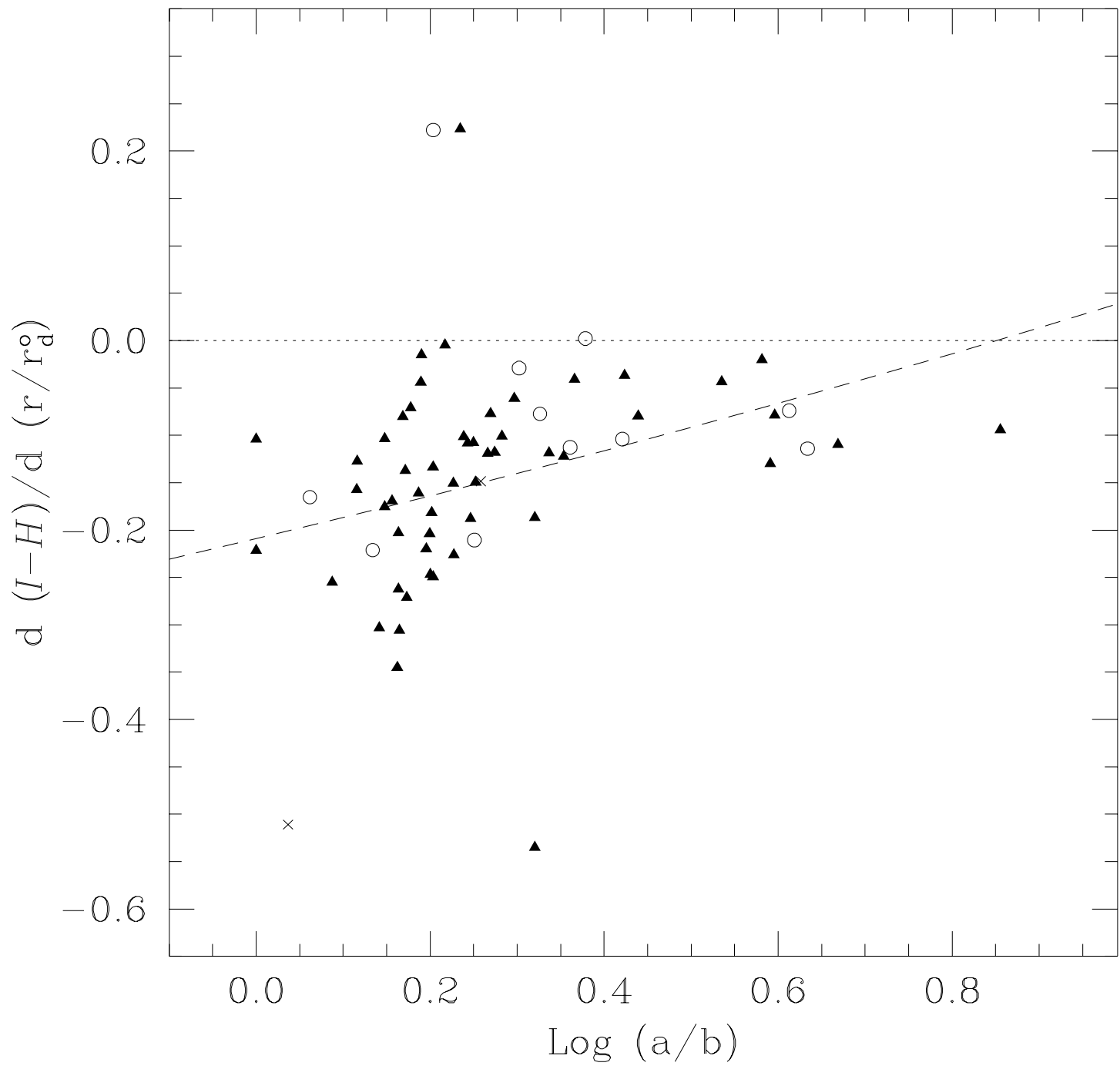
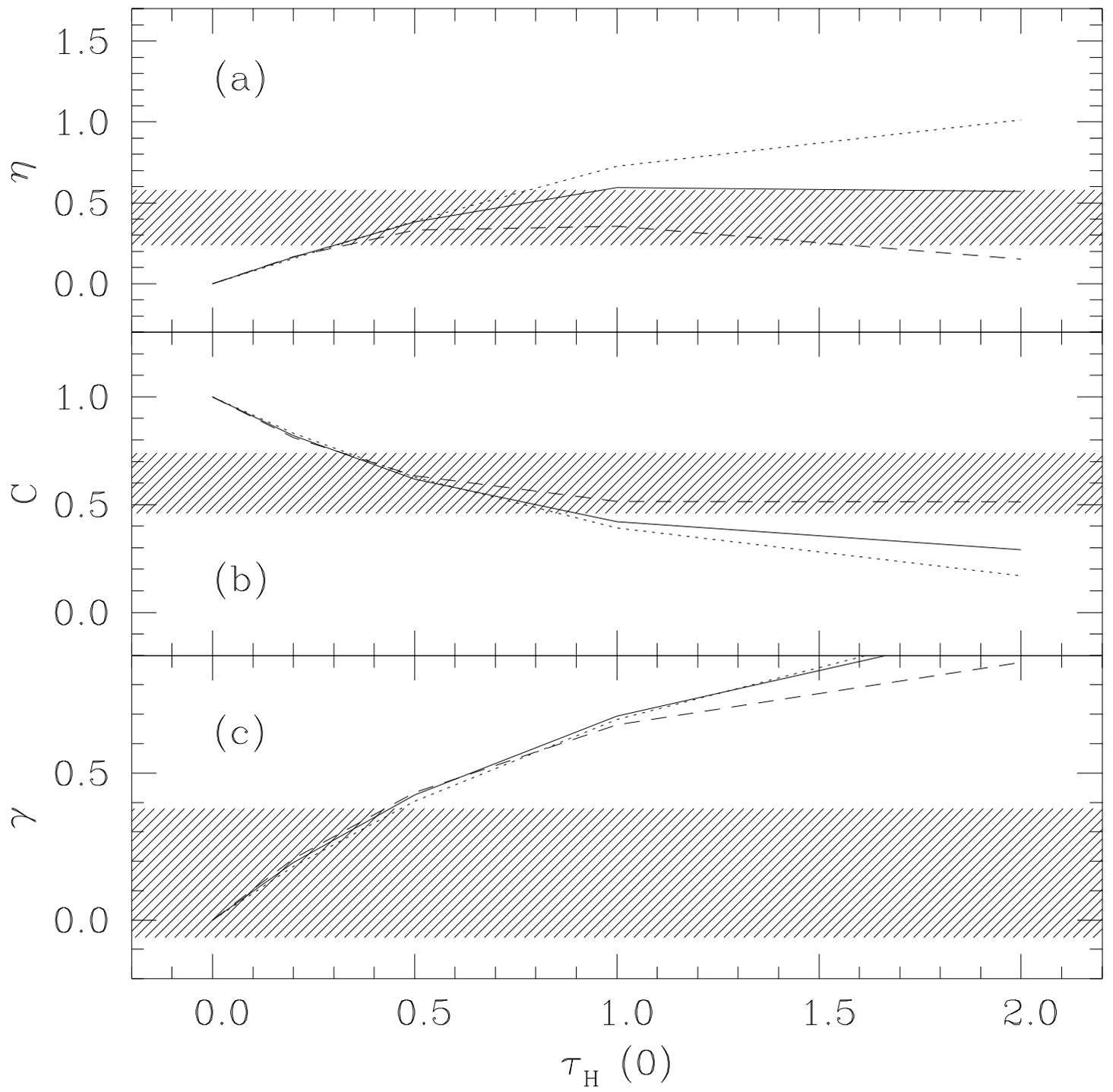
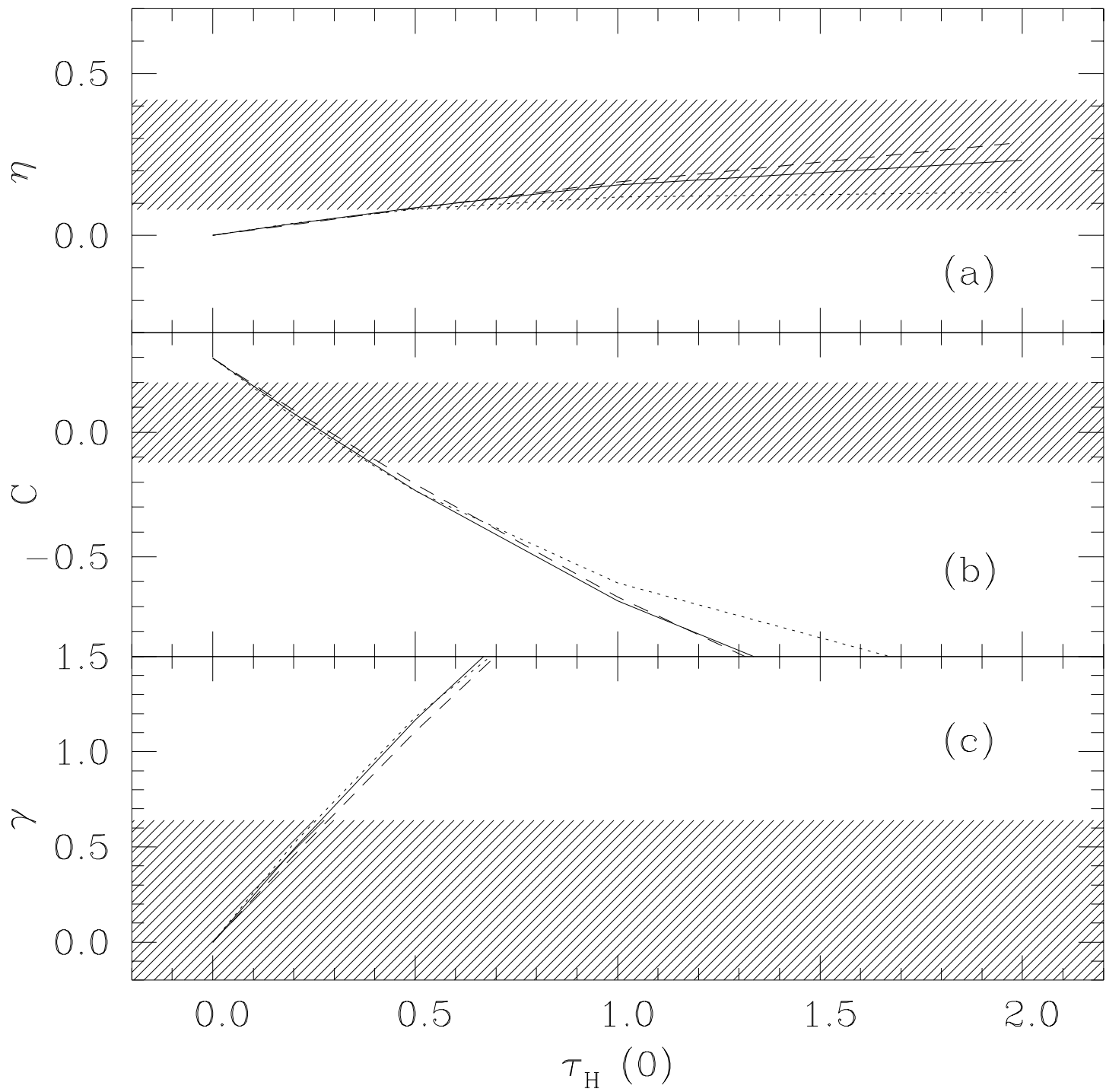
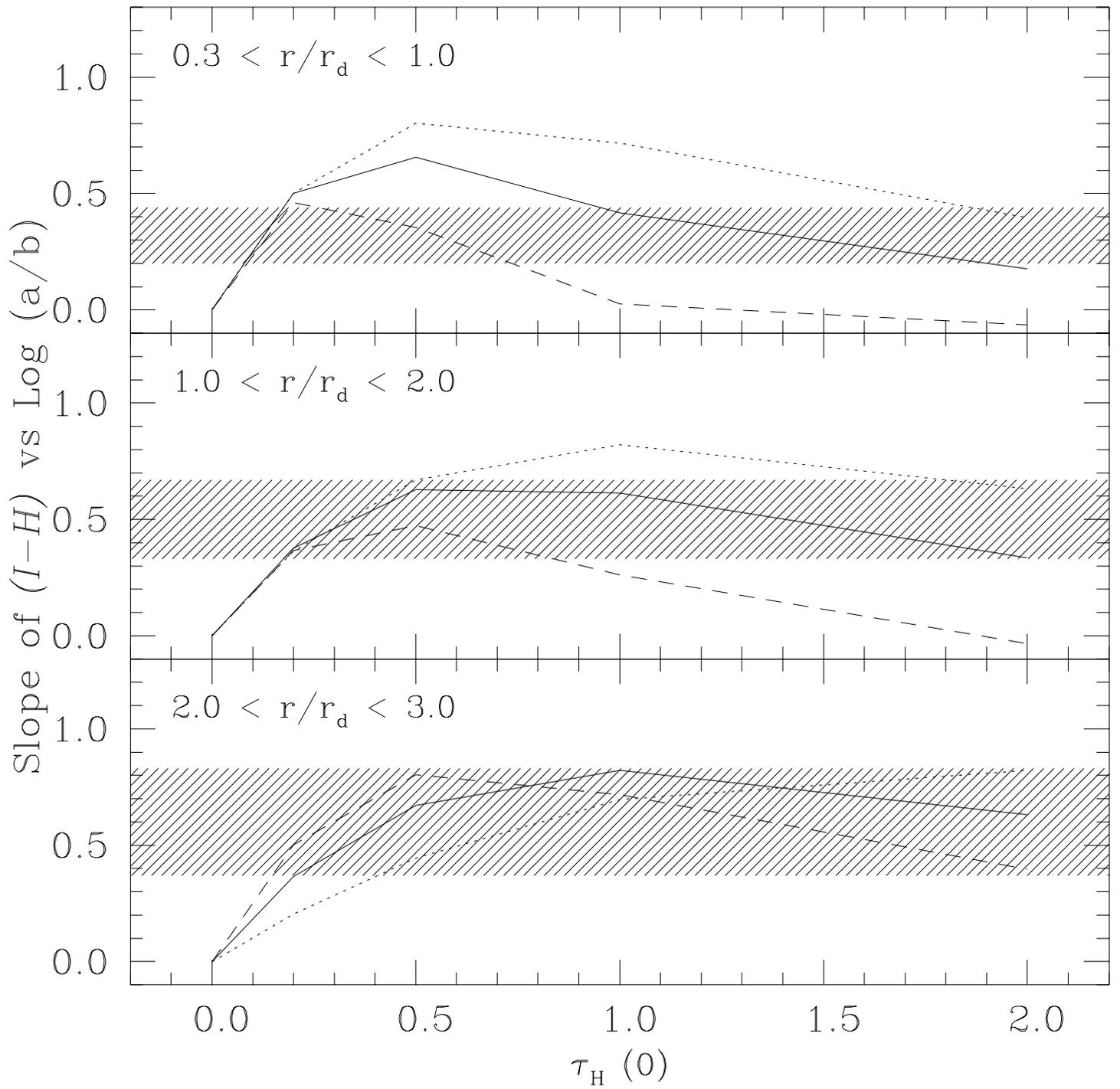
\end{document}